\documentclass[a4paper,12pt]{article}
\usepackage[labelfont=bf,font=small]{caption}

\usepackage{geometry}
\usepackage[pdftex]{graphicx}
\usepackage{url}
\usepackage{graphicx}	

\usepackage{amsmath}	
\usepackage{amssymb}	

\usepackage{pdflscape}
\usepackage[utf8]{inputenc}
\usepackage{xfrac}
\usepackage{color}
\usepackage{rotating}
\usepackage{tabularx}

\usepackage[symbol]{footmisc}

\usepackage{setspace}

\usepackage{lineno}

\usepackage{amsmath}

\usepackage[superscript,nomove]{cite}

\title{Internal mixing of rotating stars inferred from dipole gravity modes} 

\author
{May G. Pedersen$^{1,2}$\footnote{mgpedersen@kitp.ucsb.edu}, 
Conny Aerts$^{1,3,4}$, P{\'e}ter I. P{\'a}pics$^{1}$, Mathias Michielsen$^{1}$, \\
Sarah Gebruers$^{1,4}$, Tamara M. Rogers$^{5,6}$, Geert Molenberghs$^{7,8}$,\\Siemen Burssens$^{1}$,
Stefano Garcia$^{1}$, Dominic M. Bowman$^{1}$\\
\\
\normalsize{$^{1}$Institute of Astronomy, KU\,Leuven, Celestijnenlaan 200D,}\\
\normalsize{3001 Leuven, Belgium}\\
\normalsize{$^{2}$Kavli Institute for Theoretical Physics, Kohn Hall, University of California,}\\ 
\normalsize{Santa Barbara, CA 93106, USA}\\
\normalsize{$^{3}$Department of Astrophysics, IMAPP, Radboud University Nijmegen,}\\
\normalsize{P. O. Box 9010, 6500 GL Nijmegen, the Netherlands}\\
\normalsize{$^{4}$Max Planck Institute for Astronomy, Koenigstuhl 17, 69117 Heidelberg, Germany}\\
\normalsize{$^{5}$Department of Mathematics, Statistics and Physics, Newcastle University,}\\
\normalsize{Newcastle upon Tyne, UK}\\
\normalsize{$^{6}$Planetary Science Institute, Tucson, AZ 85721, USA}\\
\normalsize{$^7$I-BioStat, Universiteit Hasselt, Martelarenlaan 42, B-3500 Hasselt, Belgium}\\
\normalsize{$^8$I-BioStat, KU Leuven, Kapucijnenvoer 35, B-3000 Leuven, Belgium}\\
}

\usepackage{xcolor}

\date{}
\begin{document}
\newpage
\setcounter{page}{1}
\resetlinenumber[1]

\maketitle

\noindent\textbf{During most of their life, stars fuse hydrogen into helium in their cores. Mixing of chemical elements in the radiative envelope of stars with a convective core is able to replenish the core with extra fuel. If effective, such deep mixing allows stars to live longer and change their evolutionary path. Yet, internal mixing remained unconstrained by in-situ observations. Gravity modes probe the deep stellar interior near the convective core and allow us to calibrate internal mixing processes. Here we provide core-to-surface mixing profiles inferred from observed dipole gravity modes in 26 rotating stars with masses between 3 and 10 solar masses. We find a wide range of internal mixing levels across the sample. Stellar models with stratified mixing profiles in the envelope reveal the best asteroseismic performance. Our results provide observational guidance for 3-dimensional hydrodynamical simulations of transport processes in the deep interiors of stars.
}\\

\noindent Stars more than twice as massive as the Sun perform the hydrogen fusion in their core via the CNO-cycle, where isotopes of carbon, nitrogen, and oxygen act as catalysts in the nuclear reactions \cite{Nomoto2013}. The large amount of energy released in the CNO-cycle causes the cores of these stars to be convective and fully mixed on a dynamical time scale. As a consequence all the hydrogen that enters the convective core can be used as fuel.  For this reason, any mixing processes occurring in the transition region between the convective core and the envelope and managing to transport chemical elements into the core have a major effect on the evolution of all stars born with a convective core \cite{MaederMeynet2000}.

The global distribution of the chemical elements inside a star is determined by numerous dynamical processes, aside from nuclear reactions. The transport of elements caused by gradients of physical quantities is a diffusive process \cite{Salaris2017}, while global large-scale displacements, such as circulation due to rotation, happen via advection \cite{Georgy2013}. Numerous transport processes with a diversity of efficiencies, interactions, and time scales act together in the radiative envelope \cite{Zahn1992,Chaboyer1995,Mathis2005,Rogers2013} and have the potential to inject fresh hydrogen into the convective core, leading to a more massive helium core as long as the hydrogen fusion continues.  Conversely, material processed by core fusion may be transported to the surface of the star, where it changes the abundances \cite{Brott2011}.

The evolution of the mass fraction of a chemical element $i$ at distance $r$ from the stellar center, $X_i(r)$, requires solving a 3-dimensional diffusion-advection transport equation, leading to latitudinal variation in the extent of mixing, as well as time-dependent mixing profiles\cite{Deupree1998}.
However, in the case of a spherical star with strong horizontal turbulence due to dynamical processes, 
it has been shown that
the vertical advection can be treated diffusively \cite{ChaboyerZahn1992},
which is the approach we adopt here. 
In this case, the transport equation simplifies to

\begin{equation}
\label{transport-eq}
\displaystyle{\frac{\partial X_i(r)}{\partial t}} = \displaystyle{{\cal R}_i (r)}
+ \displaystyle{\frac{1}{\rho (r) r^2} \frac{\partial}{\partial r} \left[\rho (r) r^2 
D_{\rm mix} (r)
\frac{\partial X_i (r)}{\partial r} \right],}
\end{equation}

\noindent where ${\cal R}_i (r)$ is the local rate of change of $X_i (r)$ due to nuclear reactions and $D_{\rm mix} (r)$ is the mixing profile from the core to the surface of the star covering three regions: the convective core with mixing coefficient $D_{\rm conv} (r)$, the radiative envelope with coefficient $D_{\rm env} (r)$, and the core boundary layer, which is the transition zone between the two, with mixing coefficient $D_{\rm cbl} (r)$.

Extensive theoretical and numerical computations of transport processes for core boundary layers \cite{Zahn1991,Freytag1996} and stellar envelopes \cite{Pinsonneault1997,Charbonnel2010} have been made and their results included in stellar models across a large mass range \cite{Dotter2008,Georgy2013}. Surface abundances and model-independent dynamical masses of massive eclipsing binaries have been used to evaluate transport processes \cite{Morel2008,Brott2011,Martins2012,Tkachenko2020}. These observational studies reveal that $D_{\rm mix} (r)$ is the dominant uncertainty in the theory of stellar evolution for the majority of single and binary stars born with $M\gtrsim\,1.2\,$M$_\odot$.

Asteroseismology is a powerful tool making stellar interiors accessible to observational probing \cite{Aerts2010}. It relies on the interpretation of detected oscillation modes, which are sensitive to the local conditions in the deep stellar interior. Gravity (g) modes are particularly sensitive to the physics of the convective boundary layers and are therefore an excellent {\it in-situ} probe to assess $D_{\rm cbl} (r)$. A key observational diagnostic is the period spacing pattern of g~modes with the same degree $\ell$ and azimuthal order $m$, but with consecutive radial orders $n$, $\Delta P_{\ell,m,n} \equiv P_{\ell,m,n} - P_{\ell,m,n-1}$ \cite{Miglio2008}. This diagnostic quantity has been used to probe the physics in convective boundary layers of evolved stars and white dwarfs with short-period g~modes (periods of minutes to hours) and slow rotation (periods of days) \cite{Bossini2015}.
Here, we measure period spacing patterns of 26 slowly pulsating B-type (SPB) stars, whose g~modes and rotation have similar periods (of the order of a day), with the aim to infer $D_{\rm mix}(r)$ throughout their interior. Their photometric light curves assembled by the {\it Kepler\/} space telescope \cite{Koch2010} shown in Figure~1 (and Supplementary Figures\,1 to 26) are subjected to Fourier analysis and to the method of iterative prewhitening (see {\it Methods}) to derive period spacing patterns $\Delta P_{\ell,m,n}$ of dipole modes
as in Figure~2 (and Supplementary Figures\,1 to~26). 
For a chemically homogeneous, non-rotating star the period spacing patterns are constant and their values are mainly determined by the mass and age of the star, with more massive and younger stars having higher period spacing values. Deviations from constant patterns result due to changes in the internal chemical profiles, $X_i(r)$, occurring naturally as the star evolves. The patterns are further modified by internal mixing processes. The age and internal mixing profile, $D_{\rm mix} (r)$, hence determine the overall morphology of the patterns. The rotation of the star induces a slope in its period spacing patterns. This slope increases for higher rotation rates and shifts the patterns towards shorter periods in the case of prograde modes. Of the six stars in Figure~2, KIC~8714886 and KIC~4936089 have the lowest rotation rates and KIC~8714886 is more massive than KIC~4936089, which explains why the period spacing values of KIC~8714886 are higher than those of KIC~4936089. The
observed patterns are used to derive an initial estimate of the rotational frequency in the core boundary layer, $\Omega_{\rm rot}$, from their slope \cite{VanReeth2016}. We find that almost all the detected dipole modes occur in the gravito-inertial regime,
where the mode frequencies are below twice the rotation frequency.
This regime requires pulsation computations to be done from
a non-perturbative treatment of the Coriolis acceleration \cite{Bouabid2013}.

Modelling of the dipole period spacings is performed using a grid-based statistical approach allowing for uncertainty in the theoretical period spacing predictions due to imperfect input physics of the equilibrium models. We consider eight grids of non-rotating 1-D equilibrium models and compute their oscillation modes in the presence of the Coriolis acceleration\cite{Aerts2018}. The stellar models are represented by a set of fixed input physics $\boldsymbol{\psi}$ (see {\it Methods}) and a number of free parameters $\boldsymbol{\theta}$ discussed below. Each grid covers the entire phase of hydrogen fusion in the core, the mass range $M_{\rm ini}\in [2.75,10.0]\,$M$_\odot$ and initial metallicity $Z\in [0.003, 0.04]$, but having a different mixing profile in the core boundary layer $D_{\rm cbl} (r)$ and in the envelope $D_{\rm env} (r)$. Each of these two profiles has one free parameter: $\alpha_{\rm cbl}$ and $D_{\rm env, 0}$, respectively. Here, $\alpha_{\rm cbl}$ is a length scale connected with the size of the core boundary layer, while $D_{\rm env, 0}$ represents the level of mixing at the bottom of the radiative envelope (see {\it Methods}).
For $D_{\rm cbl} (r)$ the profile due to either convective penetration \cite{Zahn1991} or diffusive exponential overshooting \cite{Herwig2000} is adopted. 
For the envelope, a multitude of mixing profiles caused by various physical phenomena occurs in the literature. Here, we utilise four typical profiles, $D_{\rm env} (r)$: constant, wave mixing \cite{Rogers2017}, mixing due to vertical shear resulting from instabilities \cite{MathisPalacios2004}, or meridional circulation combined with large horizontal and vertical shear \cite{Georgy2013},
all of which the effect of mixing can be described diffusively. The resulting eight different $D_{\rm mix} (r)$ are illustrated in Figure~3 and represented by $\boldsymbol{\psi}_1, \dots, \boldsymbol{\psi}_8$ as indicated in each of the subpanels of the figure.
For each of these eight grids we compute statistical models which predict the theoretical period spacing values as a function of the components of $\boldsymbol{\theta}$, allowing us to refine the grid resolution between each of the grid points without having to compute additional stellar models and their oscillation properties (see {\it Methods}).
With our approach, we provide an asteroseismic evaluation of mixing based on a sample of g-mode pulsators treated in a homogeneous way, rather than just treating one star at a time as done so far \cite{Moravveji2016}. 
Since the mixing profiles are expected to change during the evolution but it is unknown in what way \cite{Deupree1998}, we evaluate whether $\alpha_{\rm cbl}$ or $D_{\rm env, 0}$ are associated with the evolutionary stage of the SPB stars in our sample.

Asteroseismic modelling of the 26 SPB stars based on their gravito-inertial dipole period spacings (cf.\,Figure~2 and Supplementary Figures 1 to 26), 
delimiting the permitted parameter space to that denoted by the spectroscopic and astrometric constraints for each star,
is done by maximum likelihood estimation of the six free parameters upon which each of the eight model grids are built (see {\it Methods}). This leads to each star's mass, metal mass fraction, evolutionary status, interior rotation frequency, convective boundary mixing, and mixing at the bottom of the envelope, represented by the parameter vector ${\boldsymbol{\theta}}\equiv\,(M_{\rm ini},Z,X_{\rm c}/X_{\rm ini},\Omega_{\rm  rot},\alpha_{\rm cbl},D_{\rm env, 0})$. Here, $X_c$ is the fractional mass of hydrogen left in the fully mixed convective core and is a proxy for the stellar age, while $X_{\rm ini}$ is the initial hydrogen mass fraction. Figure~2 and Supplementary Figures\,1 to 26 show the theoretical period spacings for the grid that best represents this measured diagnostic for each of the 26 SPB stars (the accompanying ${\boldsymbol{\theta}}$ is listed in Supplementary Table~1). 
So far, only two of these SPBs were asteroseismically modelled with non-perturbative inclusion of the Coriolis acceleration; in both cases a constant level of envelope mixing was enforced and precision estimation of ${\bf\theta}$ was not considered \cite{Moravveji2016,Szewczuk2018}.
We assess the 6-D uncertainty regions of ${\boldsymbol{\theta}}$ from a Monte Carlo approach (see {\it Methods}) and compute a weighted average for the stellar mass, evolutionary stage, metallicity and rotation frequency across the eight grids (Supplementary Table~2).

Each of our stars' dominant frequency extends the range of this observable obtained previously for 13 high-amplitude SPB stars studied from ground-based data\cite{DeCatAerts2002}
(see Supplementary Figures~27 and~28 and Supplementary Information).
To quantify whether these dominant frequencies and their amplitudes correlate with the effective temperature, surface gravity, luminosity, and stellar mass, we calculate Spearman's rank correlation coefficients $r_s$, which take values between $-1$ and $+1$, where $r_s = +1$ indicates a perfect positive correlation,  $r_s = -1$ a perfect negative correlation, and $r_s = 0$ uncorrelated data. The $r_s$ values are listed in Supplementary Table~3 for the cases where the $p$-values are $< 0.05$, implying that we can reject the null-hypothesis of them being equal to zero at the 95\% confidence level. We
find that the amplitudes of the dominant modes are negatively correlated with the mass, effective temperature, and luminosity of the star and positively correlated with the surface gravity for the 26~SPB stars. 
Our sample covers the entire SPB instability strip \cite{Szewczuk2017} and rotation rates from almost zero up to almost the critical rotation rate, i.e., $\Omega_{\rm rot}/\Omega_{\rm crit}\in [0,1]$. 
The asteroseismic results reveal Gaia DR2 luminosities and spectroscopic masses of B stars to be underestimated, which is a result of the lower effective temperature estimates from the Gaia DR2 data\cite{Anders2019,Pedersen2020} (Supplementary Information, Supplementary Figure~30 and Table~5).

None of the mixing profiles provides the best solution for all 26 SPB stars, i.e., diversity occurs in the internal mixing properties, with a wide variety of mixing efficiencies. We find that the majority of stars, i.e., 17 of the 26 SPB stars, are best modelled via convective penetration in the core boundary layer, while diffusive core overshooting offers a better explanation for the other nine. Eleven SPB stars reveal the best match for the envelope mixing profile based on vertical shear mixing, while seven
stars are best modelled with a gravity-wave mixing profile, five
with a constant profile, and four with a profile combining meridional circulation with vertical shear (cf. Figure~4 and Supplementary Information). 
Figure~5
visualizes the inferred mixing profiles of the optimal solutions found for all 26 SPB stars as listed in Supplementary Table~1 and reveals a wide range of mixing levels across the evolution, with $D_{\rm env,0} \in [12,8.7\times 10^5]\,$cm$^2$\,s$^{-1}$. We find that nine out of 26 SPB stars rotate above 70\% of their critical break-up velocity, which implies that future modelling based on 2-D stellar evolution models should be attempted whenever proper tools become available.

Figure~6 shows correlations among estimated parameters and two inferred quantities of importance for the further evolution of the stars resulting from $\boldsymbol{\theta}$, i.e., the fractional Schwarzschild convective core mass and size, $m_{\rm cc}/M$ and $r_{\rm cc}/R$. 
Despite several large and asymmetrical uncertainty intervals resulting from projections of the 6-D elongated uncertainty regions onto 1-D, we find that both of these quantities decrease as the evolution progresses, as expected from theory\cite{Kippenhahn2012} and from hydrodynamical simulations \cite{Deupree1998}.
The g~modes allow for proper inference of the core masses despite considerable uncertainty for $\alpha_{\rm cbl}$ in the 6-D fitting. The convective core mass expressed as a percentage of total stellar mass ranges from $\sim$30\% near the zero-age-main-sequence (ZAMS) and stays above $\sim$6\% near the terminal-age-main-sequence (TAMS), confirming the need for higher-than-standard core masses in eclipsing binaries \cite{Tkachenko2020} and in young open clusters \cite{LiSundeGrijs2019,Johnston2019}.
Through the calculation of Spearman's rank correlation coefficients (Supplementary Table~7) we find a strong correlation between $r_\text{cc}/R$ and $m_\text{cc}/M$ as well as between the core masses and radii and the main-sequence evolutionary stage $X_\text{c}/X_\text{ini}$, as expected from theory.
The convective core size is uncorrelated to the level of envelope mixing, while the convective core mass correlates moderately with it, as shown in Figure~6 for values $D_{\rm env,0}\gtrsim 10^3\,$cm$^2$\,s$^{-1}$. This points to envelope material getting efficiently transported to the core for such levels of mixing.

Aside from the slow rotator KIC\,8459899, whose asteroseismic modelling led to low metallicity, $D_{\rm env,0}$ increases with increasing $\Omega_{\rm rot}/\Omega_{\rm crit}$ irrespective of the mass and evolutionary stage.
Higher $D_{\rm env,0}$ values lead to higher core masses (Supplementary Table~7), while no correlation is found between $D_{\rm env,0}$ and $X_{\rm c}/X_{\rm ini}$ at the 95\% confidence level. For KIC\,3240411, similar results to ours
were achieved from modelling based on 1-D equilibrium models with rotational mixing\cite{Szewczuk2018}. We added published results for two slowly rotating SPB stars in Figure~6. These were modelled using the same input physics as in our $\psi_1$\cite{Wu2019,Wu2020}, but by adopting a perturbative treatment for the Coriolis acceleration rather than the TAR as in the current work. For reasons of consistency, these two stars were not included in our computations of the Spearman's rank correlation coefficients 
listed in Supplementary Table~7. Figure~6 shows that the three slowest rotators, where two are metal-poor and one is metal-rich, reveal considerable levels of envelope mixing.  
We find no other clear correlations between the remaining estimated ${\boldsymbol{\theta}}$ values and hence do not provide the correlation coefficients for them.

Our homogeneous analysis based on a sample of g-mode pulsators 
offers the opportunity to evaluate the quality of the input physics of stellar models in the covered mass and age range. We conclude that the internal mixing profiles of almost all SPB stars
as inferred from asteroseismology
are radially stratified instead of constant. Future derivation of rotation and mixing profiles, $\Omega (r,t)$ and $D_{\rm mix}(r,t)$, without having to rely on predefined time-independent profiles as done so far, can be achieved from a much larger sample of SPB stars with sufficient identified g~modes having proper probing capacity. Such observationally calibrated mixing profiles and the resulting helium core masses near core hydrogen exhaustion
constitute important asteroseismic input to improve stellar evolution models and chemical yield computations for the evolved stages of stars born with a mass above three solar masses.


\newpage

\newpage
\begin{center}
\includegraphics[width=\linewidth]{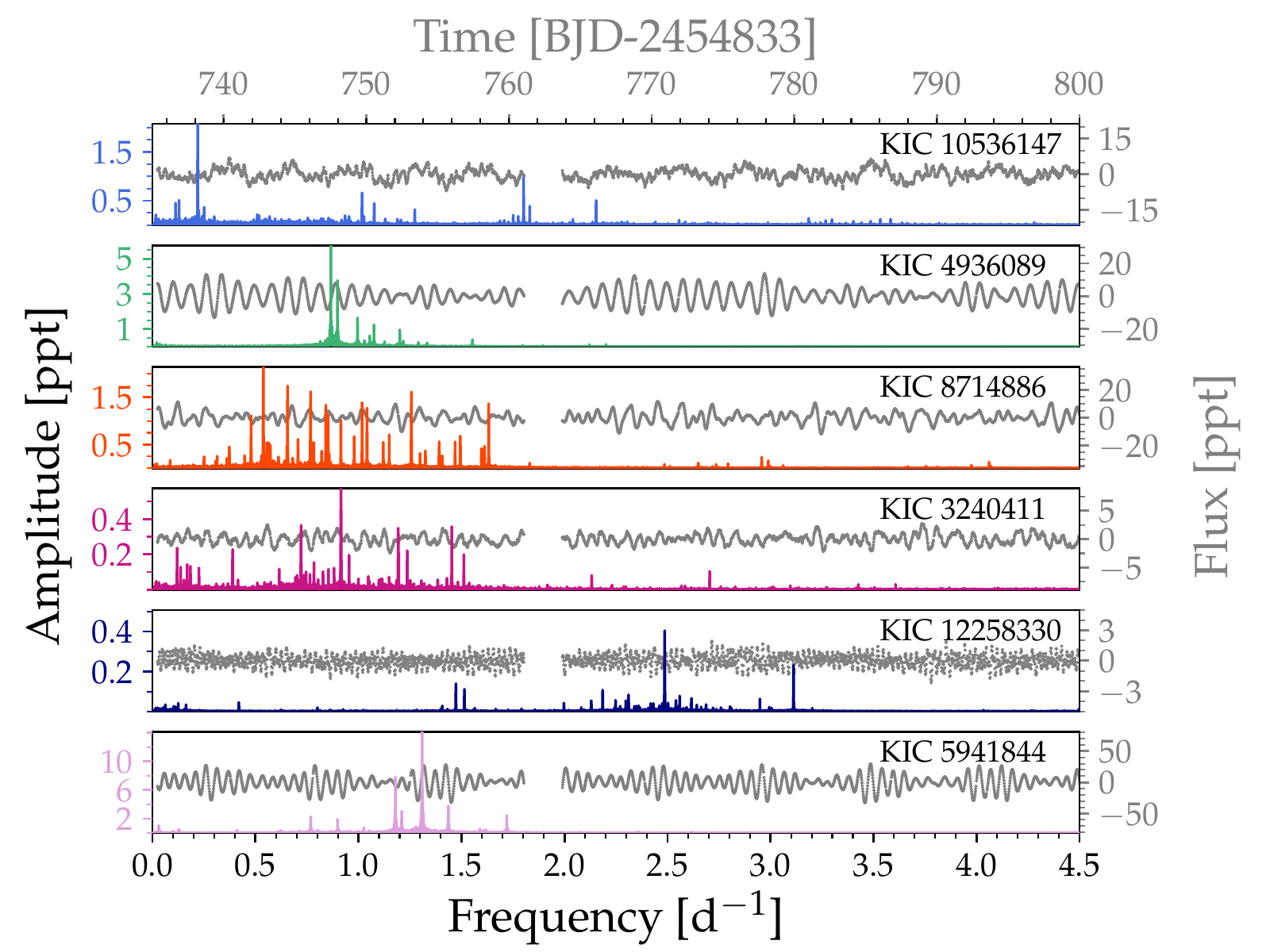}
\end{center}

\noindent \textbf{Figure 1:} {\bf Light curves overplotted with amplitude spectra of six Slowly Pulsating B stars.}
Excerpts from {\it Kepler\/} long-cadence ($\sim$30 minutes per point) light curves (flux as a function of time in grey dots) of six new SPB stars whose {\it Kepler\/} Input Catalogue (KIC) identification is indicated. The oscillation spectra (amplitude as a function of frequency, coloured lines) derived from the full light curves with a total duration of $\sim$1500\,days are overplotted and reveal multiple gravity-mode frequencies with periods of the order of days. \\

\newpage
\begin{center}
\includegraphics[width=\linewidth]{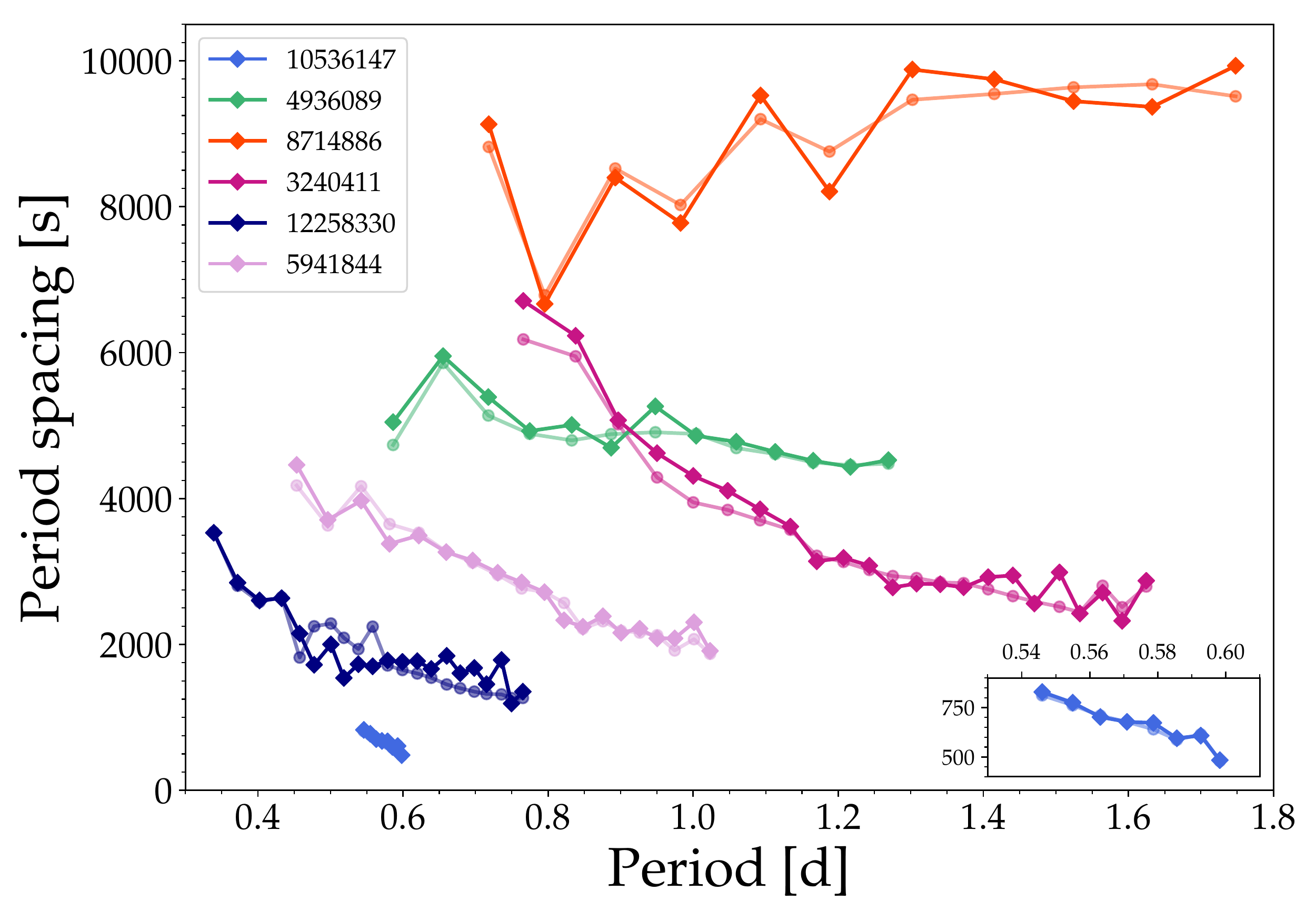}
\end{center}

\noindent \textbf{Figure 2:} {\bf Gravity-mode period spacing patterns of six Slowly Pulsating B stars.}
Observed dipole-mode period spacings, $\Delta P_{1,m,n}$ (indicated in coloured diamonds) of the six SPB stars whose light curves and amplitude spectra are shown in Figure~1 are compared with the theoretically predicted values (bullets in the same colour with lighter colour tone) based on the best stellar evolution model from eight model grids. The formal errors of the observed values are smaller than the plotted symbol size for most of the detected modes (see {\it Methods}).
The inset contains a zoom in on the SPB with the lowest period spacings in the sample. 
The slope of the patterns correlates with the near-core rotation rate. Younger stars and stars of higher mass result in higher period spacing values, while the pattern morphology is mainly determined by the evolutionary stage and internal mixing.\\

\newpage
\begin{center}
\includegraphics[width=\linewidth]{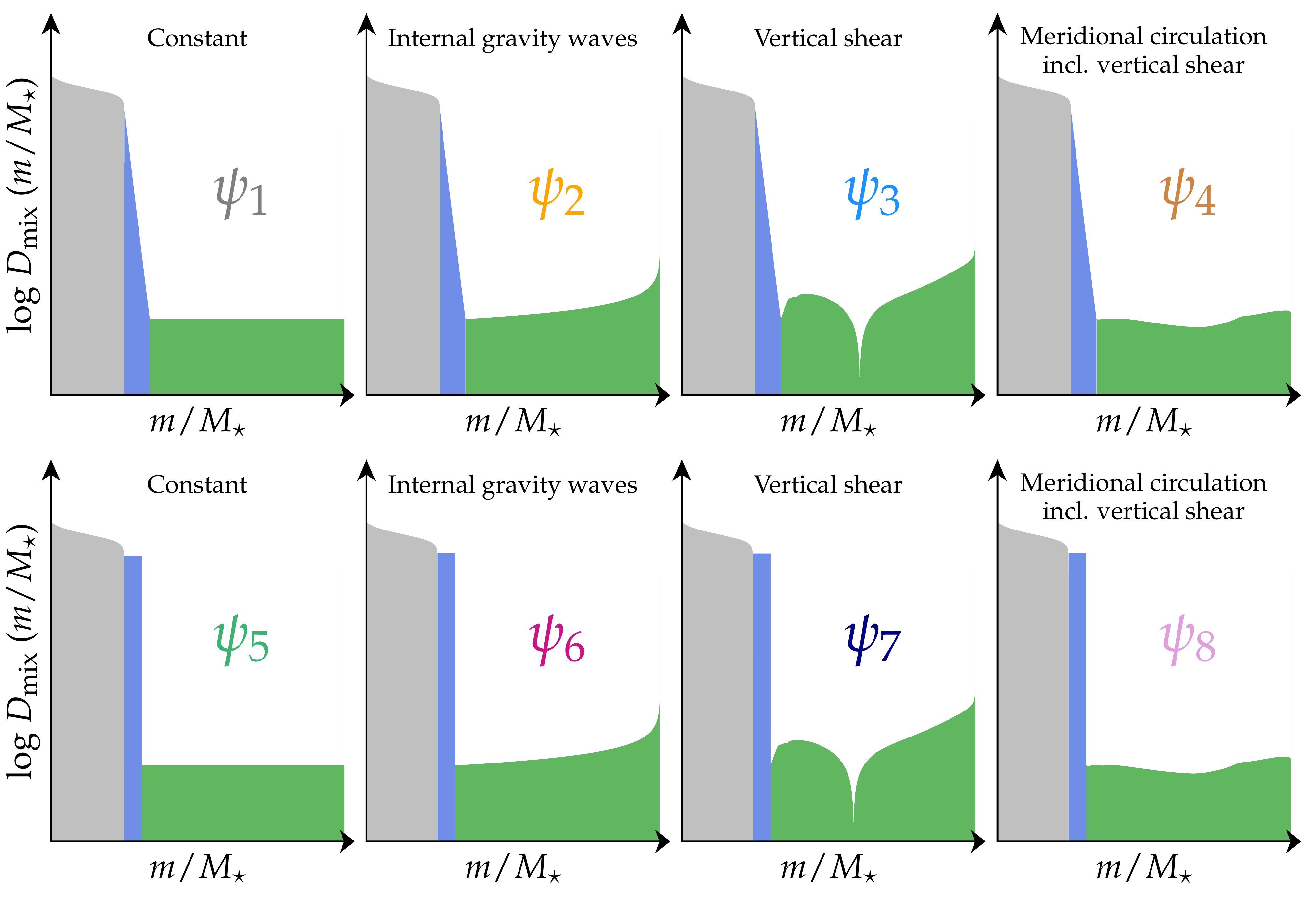}
\end{center}

\noindent \textbf{Figure 3:} {\bf Schematic 
representation of the considered mixing profiles.}
The convective core, convective boundary, and envelope mixing levels as a function of fractional mass inside the stellar models are indicated in grey, blue, and green, respectively. Upper panels: diffusive exponentially decaying overshooting\cite{Herwig2000}, lower panels: convective penetration\cite{Zahn1991}. From left to right: constant envelope mixing, mixing caused by internal gravity waves\cite{Rogers2017}, mixing due to vertical shear connected with instabilities\cite{MathisPalacios2004}, and due to meridional circulation caused by rotation combined with vertical shear\cite{Georgy2013}.\\

\newpage
\begin{center}
\includegraphics[width=\linewidth]{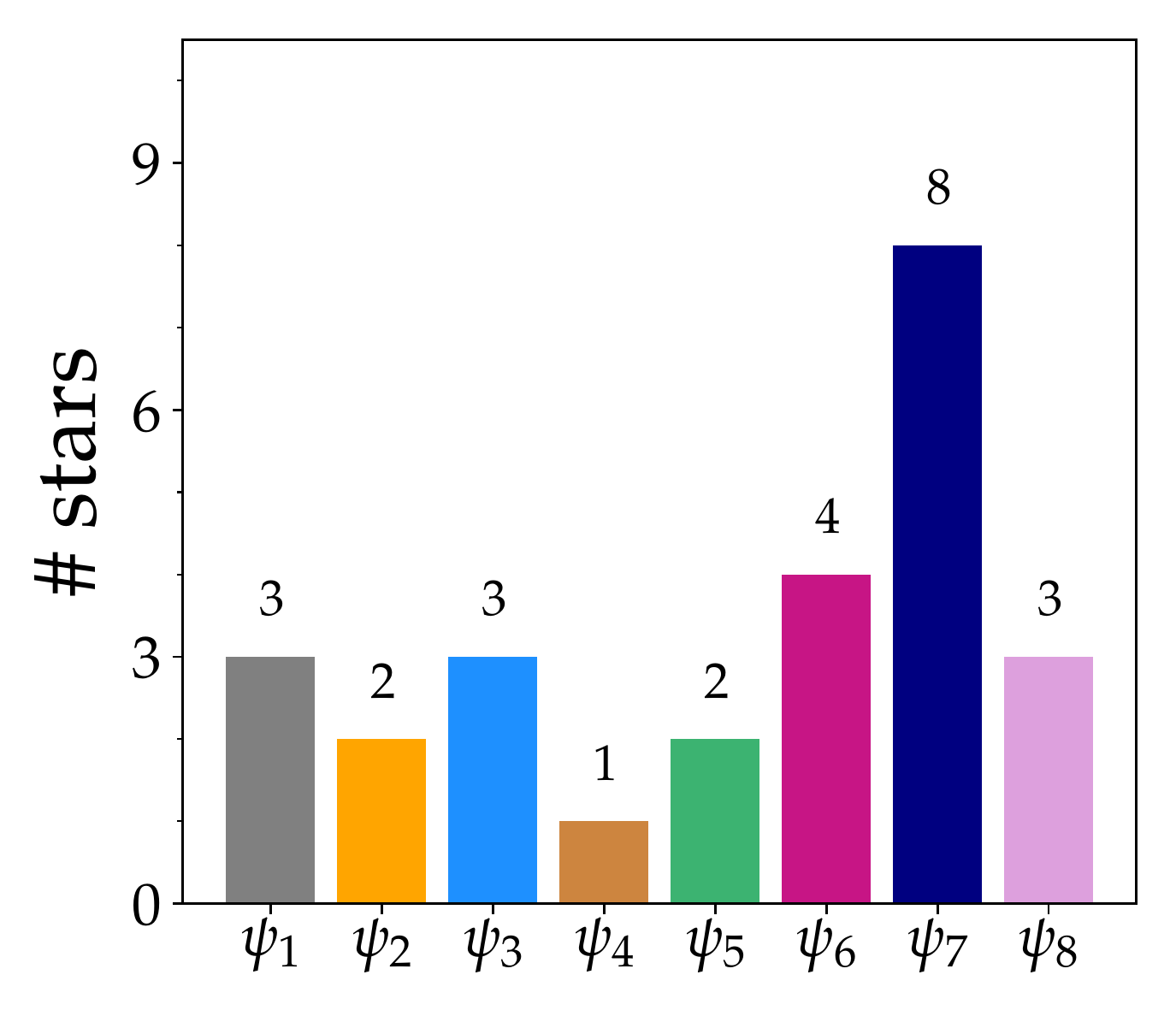}
\end{center}

\noindent \textbf{Figure 4:}
{\bf Population of the eight model grids in terms of model capacity.}
The 26 SPB stars are distributed over the eight stellar evolution model grids according to the best solution. The colour of the bars corresponds to the colour of the symbols in Figure~3.\\

\newpage
\begin{center}
\includegraphics[width=\linewidth]{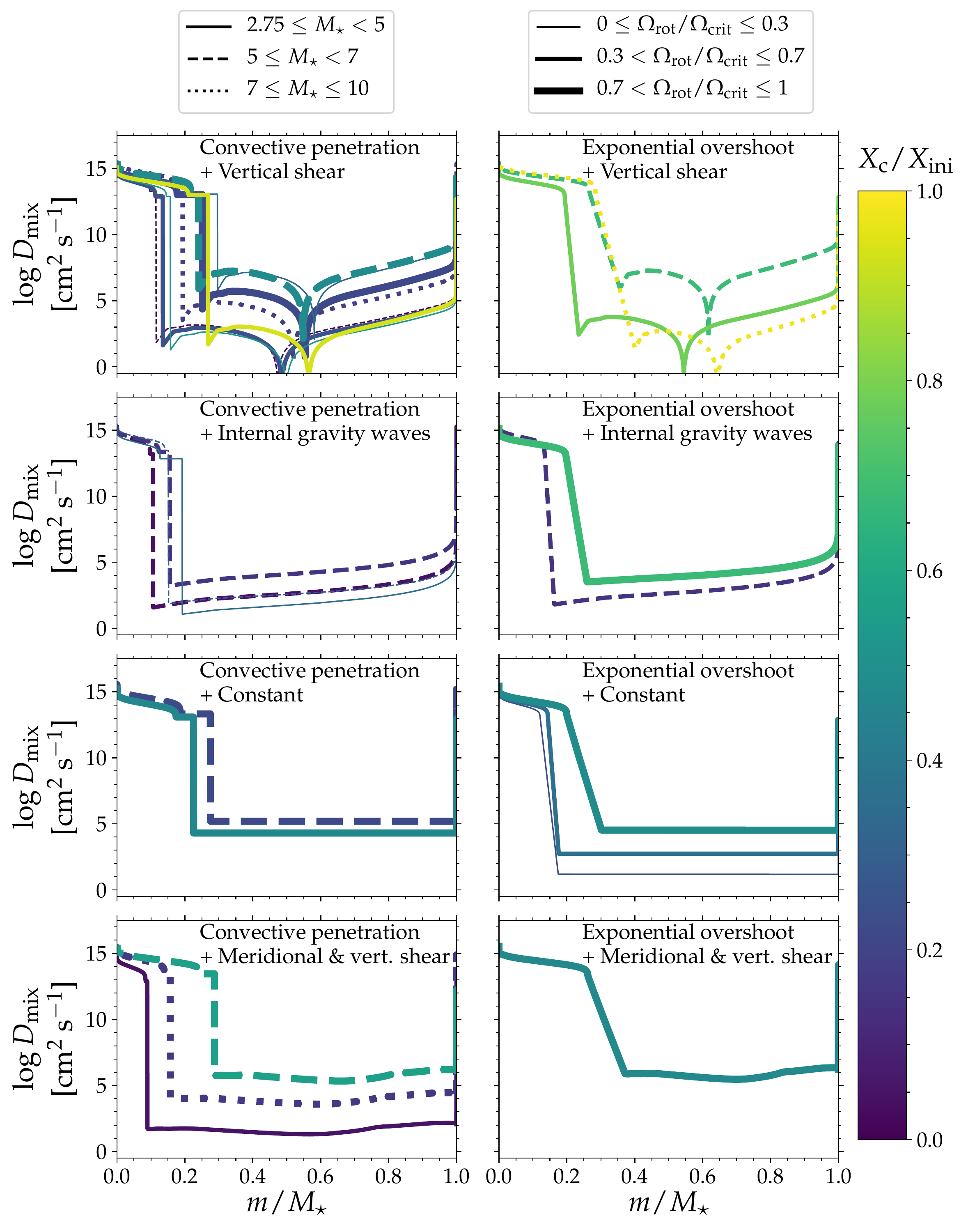}
\end{center}

\noindent \textbf{Figure 5}: {\bf Inferred internal mixing profiles for 26 
Slowly Pulsating B stars.}
The results from the asteroseismic modelling based on the detected gravito-inertial dipole modes are overplotted for the 26 best fitting models. The individual profiles are colour coded according to their main-sequence evolutionary stage $X_\text{c}/X_\text{ini}$, while the linestyle and thickness are related to the mass and rotation rate, respectively.\\

\newpage
\begin{center}
\includegraphics[width=\linewidth]{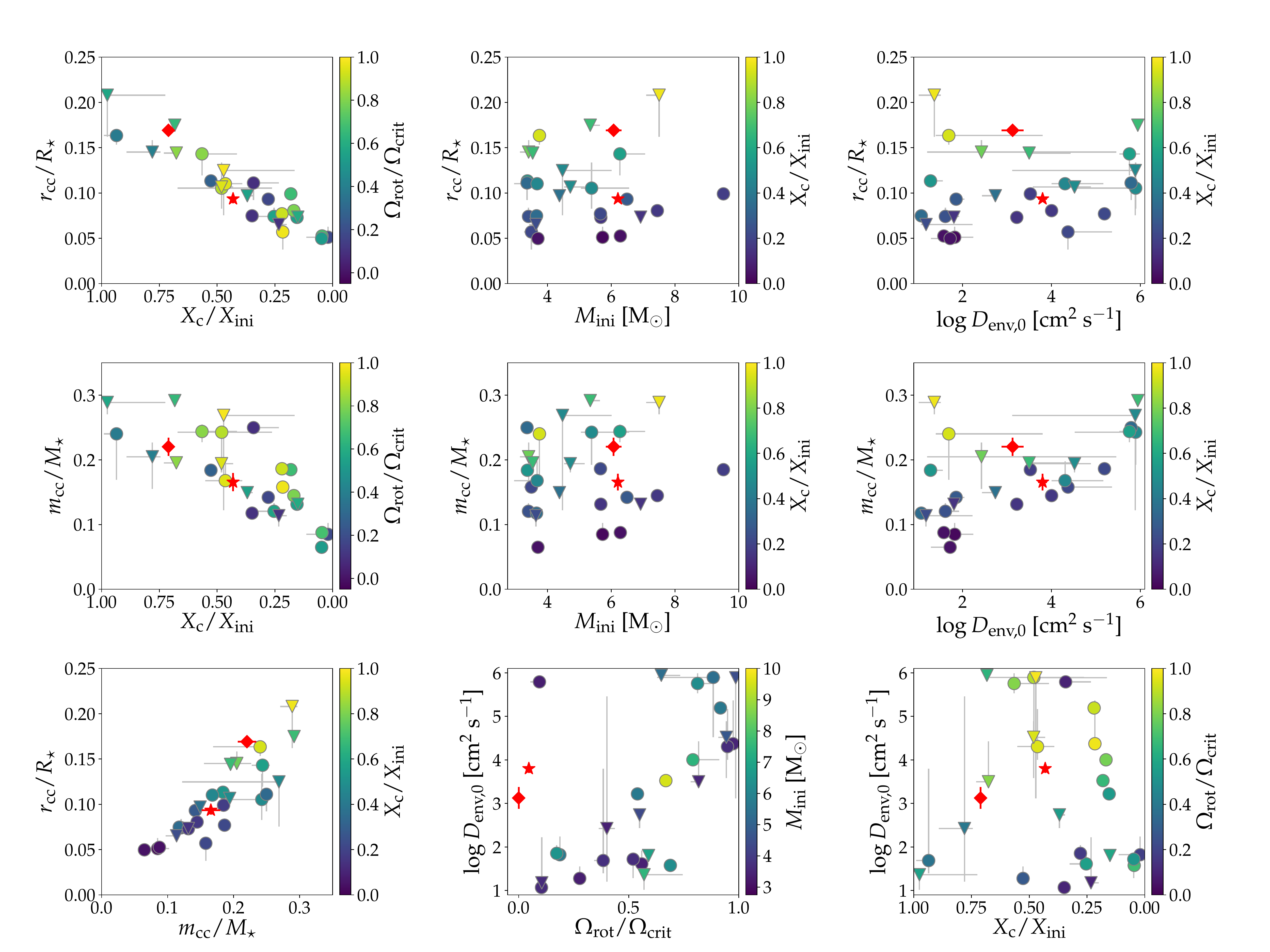}
\end{center}

\noindent \textbf{Figure 6:} {\bf Correlations among estimated parameters and 
inferred quantities for the sample.}
Stars best modelled by diffusive exponential overshooting in the core boundary layer are indicated with triangles, while those best fit by convective penetration are shown as circles. Projections of the 6-D uncertainty regions in 1-D for the corresponding parameter
are indicated in grey. We also show published results for the two slowly rotating SPB stars KIC~8324482\cite{Wu2020} (red diamond) and HD~50230\cite{Wu2019} (red star), which have been modelled by relying on similar input physics as in our grid $\psi_1$, but by adopting a perturbative treatment of the Coriolis acceleration rather than the TAR for the pulsation computations.

\newpage

\section*{Methods}

\subsection*{Sample selection, period spacing patterns, and modelling strategy}

We selected 60 candidate Slowly Pulsating B (SPB) stars from various published {\it Kepler\/} variability catalogues, having long cadence (30-min) light curves of 4-year duration from the nominal {\it Kepler\/} mission \cite{Koch2010} and revealing at least three independent frequencies in the g-mode regime.  
This is half of the discovered SPB stars from the nominal mission and about 9\% of the monitored B-type stars\cite{Balona2015}. This restriction of excluding mono- and biperiodic g-mode pulsators was built in since our aim is to perform asteroseismic modelling based on multiple modes whose degree can be identified from period spacing patterns.
We transformed the raw {\it Kepler\/} pixel data of all the quarters for these 60 SPB candidates into light curves using star-dependent customized pixel masks. The merged (over all 18 quarters) and detrended light curves were subjected to frequency analysis by iterative prewhitening \cite{Aerts2010}. We retained frequencies with an amplitude above four times the local noise level computed over a frequency range of 1\,d$^{-1}$ centred around the considered frequency. Furthermore, we rejected frequencies that are within $2.5/T$ of one another, where $T$ is the total length of the time series, and keep only the frequency with the highest amplitude within this interval. Among all the remaining significant frequencies, harmonics and combination frequencies were identified, taking into account the frequency resolution of the 4-year time series, and excluded for the asteroseismic modelling. Further, we retained only those frequencies with amplitudes significant at $1\sigma$ level from linear regression performed in the time domain at each stage of the prewhitening, as an acceptable procedure to find period spacing patterns of maximal extent in consecutive radial order to perform the asteroseismic modelling.
This led to lists ranging from 37 to 109 independent mode frequencies to work with for 
each of
the 26 stars. Period spacing patterns were then searched for among these remaining independent frequencies\cite{Papics2017}.
In the cases that a combination frequency fits into the pattern, it was included as part of the pattern as a combination may occur by chance. We identified period spacing patterns
for 26 out of the 60 stars.
Overall, this implies that $\sim\,4\%$ of all the B stars in the nominal {\it Kepler\/} field of view revealed g-mode pulsations suitable for asteroseismic modelling.

Excerpts of the light curves, Fourier transforms of the full 4-year light curves, 
all frequencies with significant amplitudes in the light curves,
and dipole mode period spacing patterns are shown graphically for all 26 SPB stars in Supplementary Figures\,1 to 26. Overall, we treat 388 g~modes in the 26 stars, with mode frequencies ranging from 0.3525 to 3.4385\,d$^{-1}$ and amplitudes ranging from 0.0022 to 14 parts-per-thousand (ppt). The errors of the period spacings used in this work are based on the frequency errors derived from least-squares harmonic fits to the light curves at each stage of the prewhitening. These formal errors range up to $\sim\,50\,$seconds, which are much smaller than the uncertainties of the theoretical predictions for g-mode period spacings based on present-day stellar models \cite{Aerts2018}. The near-core rotational frequency of the stars was determined from the slope of the period spacing patterns\cite{VanReeth2016}.

Asteroseismic modelling was so far done for only four {\it Kepler\/} SPB stars: KIC\,10526294 \cite{Moravveji2015}, KIC\,7760689 \cite{Moravveji2016}, KIC\,3240411 \cite{Szewczuk2018}, and KIC\,8324482 \cite{Wu2020},
where the latter star is a very slow rotator with a high level of envelope mixing, interpreted in terms of shear mixing due to differential rotation by the authors. KIC\,10526294, on the other hand, is also an ultraslow rotator\cite{Triana2015} with modest envelope mixing. 
These four previous
applications considered constant envelope mixing and hardly assessed the quality of the input physics of stellar models, as this requires a systematic homogeneous modelling application to an ensemble of SPB stars. Here, we provide such an application based on a statistical approach\cite{Aerts2018}. While this may lead to less precise results compared to a grid-based approach dedicated to a single star, it offers a coherent framework, allowing to assess the quality of various theories of stellar evolution \cite{Aerts2020}. For the current application, we specifically evaluate the quality of stellar models with eight different internal mixing profiles (cf.\,Figure~3).

\subsection*{Fundamental parameters of the sample stars}

Aside from the {\it Kepler\/} light curves (Supplementary Figures\,1 to 26), high-resolution spectra for the 15 brightest SPB stars were assembled with the HERMES spectrograph mounted on the 1.2-m Mercator telescope\cite{Raskin2011} and with the ISIS spectrograph on the William Herschel Telescope, both located on La Palma, Spain. 
Three stars are known to be binaries with an orbital period much longer than the periods of the g~modes\cite{Papics2013}.
We performed a standard reduction of the data, following earlier analyses for some of the stars\cite{Papics2017}. After manually normalizing the spectra via spline fitting, we determined the effective temperature ($T_{\rm eff}$), gravity ($\log\,g$), metallicity ([M/H]), projected rotation velocity ($v\sin\,i$), and individual abundances using the publicly available Grid Search in Stellar Parameters spectral synthesis code based on synthetic spectra resulting from LTE-based model atmospheres \cite{Tkachenko2015}. The results for the global stellar parameters are shown in blue in the histograms in Supplementary Figure~31, where we also included the luminosity of the 26 stars based on Gaia DR2 astrometry\cite{Pedersen2020}. For the 11 SPB stars without high-resolution spectroscopy, we assembled lower-precision estimates of $T_{\rm eff}$, $\log\,g$, and [M/H] from the literature. For some of the histograms in Supplementary Figure~31, we also show the distribution of the sample of 37 OB-type stars with available nitrogen abundances and rotation frequencies\cite{Aerts2014}.

The asteroseismic surface nitrogen abundance covered by the two grids with shear envelope mixing is shown graphically in Supplementary Figure~32. The asteroseismic predictions indicated on the plot are those resulting from the best ${\boldsymbol{\theta}}$ for that particular grid obtained for the 26 SPB stars. The spectroscopic measurements available for 15 SPB stars are in good agreement with the asteroseismic ones, reaching the $1\sigma$ level for 11 of the 15 SPB star and the $2\sigma$ level for the four additional ones.

\subsection*{Grids of evolutionary models and their pulsation modes}

The asteroseismic modelling of the ensemble of SPB stars relies on eight 
grids of non-rotating
stellar models constructed using the MESA code, 
adopting the MESA Equation of State \cite{Paxton2019}. We rely on 1-D spherically symmetric equilibrium models, where the effects of rotation, magnetism, 
waves, 
radiative levitation, etc.,  are only taken into account at the level of the element transport via Eq.\,(\ref{transport-eq}) by means of an unknown local time-independent
diffusion coefficient $D_{\rm mix} (r)$\cite{Aerts2020}.
This approach does not include angular momentum transport, since its theory remains uncalibrated for the phase of  hydrogen fusion according to asteroseismic measurements \cite{Aerts2019,Aerts2020}. For this reason, we do not include such transport but rather estimate the internal rotation frequency at the evolutionary stage of each star in the sample.
We use the Ledoux criterion for convection and include the predictive premixing scheme to compute the convective core boundary \cite{Paxton2019}.

OP opacity tables \cite{Seaton2005} applied to the initial chemical mixture of nearby B-type stars \cite{Przybilla2013} were used\cite{Moravveji2015}. The models were evolved starting from the Hayashi track to the end of the hydrogen fusion in the core, ensuring time steps below 0.001\% of the nuclear time scale. Once the zero-age-main-sequence (ZAMS) is reached the mesh refinement of the models near the core boundary region is increased, and the Vink hot wind scheme\cite{Vink2001} is switched on assuming a wind scaling factor of 0.3 \cite{Bjorklund2020}. The atmospheric table option in MESA is applied as the outer boundary conditions for the stellar models, and the full pp-chain and CNO cycle networks are included in the nuclear network. For a given initial $Z$, the initial hydrogen $X_\text{ini}$ and helium $Y_\text{ini}$ mass fractions are adjusted such that the ratio $X_\text{ini}/Y_\text{ini} = X_\star/Y_\star$ is constant across all stellar models, where $X_\star$ and $Y_\star$ correspond to the Galactic standard values\ for B stars in the solar neighbourhood\cite{Przybilla2013}.

The only difference in the input physics between the eight grids is the internal chemical mixing profile described by the local diffusion coefficient $D_{\rm mix} (r)$. The eight choices for the profiles of the diffusion coefficients are shown schematically in Figure~3. For all eight model grids $D_{\rm conv} (r)$ is based on the mixing-length theory of convection \cite{BohmVitense1958}. Two choices for the mixing profile in the core boundary layer, $D_{\rm cbl} (r)$, were considered. A first choice for $D_{\rm cbl} (r)$ is an exponentially decaying function described by the parameter $\alpha_{\rm cbl}=f_{\rm ov}$ and representing diffusive convective overshoot mixing \cite{Herwig2000} in a zone with the radiative temperature gradient (grids denoted as $\boldsymbol{\psi}_1, \ldots, \boldsymbol{\psi}_4$ in Figure~3).  A second choice is a step function based on convective penetration \cite{Zahn1991} leading to full instantaneous mixing over a distance expressed by the parameter $\alpha_{\rm cbl}=\alpha_{\rm pen}$.  In this case, the temperature gradient in the boundary layer is taken to be the adiabatic one (grids $\boldsymbol{\psi}_5, \ldots, \boldsymbol{\psi}_8$ in Figure~3). Each of these two $D_{\rm cbl} (r)$ profiles is stitched to four options for $D_{\rm env} (r)$ at the mixing level determined by the free parameter $D_{\rm env, 0}$. The four options for the envelope mixing represent 1) a constant profile \cite{Moravveji2015}, 2) a profile due to internal gravity waves \cite{Rogers2017}, 3) a profile typical of vertical shear due to various types of instabilities \cite{MathisPalacios2004}, and 4) a profile due to meridional circulation in the presence of vertical shear \cite{Georgy2013}.

For each of the eight grids, six free parameters are considered. These are the initial mass $M_{\rm ini} \in [2.75, 10] \ {\rm M}_\odot$, the initial metal mass fraction $Z \in [0.003, 0.04]$, where the range is chosen based on the observed metalicities, the ratio of the current to initial hydrogen mass fraction in the stellar core $X_{\rm c}/X_{\rm ini} \in [0.99, 0.02]$, the extent of the convective boundary mixing region ($f_{\rm ov} \in [0.005, 0.04]$ and $\alpha_{\rm pen} \in [0.05, 0.40]$), the level of envelope mixing at the position where the transition from convective core boundary mixing to envelope mixing happens ($D_{\rm env, 0} \in [10, 10^6] \ {\rm cm}^2 \ {\rm s}^{-1}$), and the rotational frequency of the star with respect to the critical rotation rate $\Omega_{\rm rot} = [0, 0.7]\ \Omega_{\rm crit}$. The dipole g-mode frequencies for each of the models in the grids were computed taking into account the Coriolis acceleration in the Traditional Approximation of Rotation (TAR), which offers a valid approximation for the range of rotation rates considered here. 
Indeed, it was shown that the TAR based on 1-D models performs well for dipole prograde and zonal modes for stars rotating up to $~\sim\,70\%$ of their critical break-up velocity by comparing the computed frequencies with those obtained from 2-D models deformed by the centrifugal acceleration\cite{Ouazzani2017}.
The pulsation frequencies using the TAR were computed with the GYRE pulsation code \cite{Townsend2018} for all radial orders in the range $|n| \in [1, 80]$ and constitute the theoretical input for the modelling procedure.

The sampling in parameter space for $(M_{\rm ini}, Z, f_{\rm ov}, D_{\rm env, 0})$ and $(M_{\rm ini}, Z, \alpha_{\rm pen}, D_{\rm env, 0})$ was done using a quasi-random sampling based on Sobol numbers\cite{Bellinger2016}. Sampling these two sets of four parameter ranges 2500 times is sufficient for determining the statistical models adopted in the modelling procedure. Supplementary Figure 33 illustrates what this 2500 grid point quasi-random sampling looks like for the $(M_{\rm ini}, Z, f_{\rm ov}, D_{\rm env, 0})$ set of parameters in comparison to a linear grid sampling. For each of the 2500 initial starting parameters, stellar models are computed and stored for $X_{\rm c}/X_{\rm ini} = 0.99, 0.95$ and for each 0.05 decrease in $X_{\rm c}/X_{\rm ini}$ down to 0.20. Below this value, we compute the stellar models in steps of 0.02 in order to account for the increasing occurrence of avoided crossings among the frequencies \cite{Aerts2020}. The last stellar model on the track has $X_{\rm c}/X_{\rm ini} = 0.02$. 
Each of the eight grids have 65000 equilibrium models upon which we base the asteroseismic modelling.
For each equilibrium model, we compute the dipole gravity modes in the TAR for five values of the rotational frequency $\Omega_{\rm rot} = [0, 0.1, 0.3, 0.5, 0.7]\ \Omega_{\rm crit}$, resulting in a total of 325000 different combinations of $\boldsymbol{\theta}$ for each of the $\boldsymbol{\psi}_1, \dots, \boldsymbol{\psi}_8$ grids. These eight grids are used to calculate predictions of period spacing values and statistical approximations thereof as described in detail in the following section.

\subsection*{Asteroseismic modelling per individual star and per grid}

The four observables $T_{\rm eff}$, $\log\,g$, [M/H], and $\log\,(L/L_\odot)$ were used to limit the range of evolution models considered for each star. We adopted $2 \sigma$ errors to ensure a 95\% probability that the star falls into the grid of models. The range in $\Omega_{\rm rot}$ to consider for the modelling is determined based on the observed rotational frequency range derived from the slope of the period spacing patterns\cite{VanReeth2016} in Supplementary Figures\,1 to~26. Statistical computations\cite{Aerts2018} are done to approximate the pulsation mode period spacings for each star for an additional 100000 quasi-randomly sampled grid points in the 6-D parameter space inside the observed error boxes.  Per grid, these statistical models are built from the original 325000 equilibrium models. Period spacing values $\Delta P_{\ell, m, n}$ are then predicted based on the varied parameters $\boldsymbol{\theta} = (M_{\rm ini}, Z, X_{\rm c}/X_{\rm ini}, \ f_{\rm ov}, D_{\rm env, 0}, \Omega_{\rm rot})$ or $\boldsymbol{\theta} = (M_{\rm ini}, Z, X_{\rm c}/X_{\rm ini}, \alpha_{\rm pen}, D_{\rm env, 0}, \Omega_{\rm rot})$.

Following hare-and-hound tests, the asteroseismic modelling is done from mode period spacing values, because this diagnostic reveals the best performance among the three tested cases of using 1) mode frequencies, 2) mode periods, and 3) mode period spacings.  The observed period spacing values $\Delta P_{1, m, n}$ are least prone to systematic uncertainties due to limitations in the input physics of the equilibrium models. The statistical models to predict the period spacing patterns are based upon a multivariate regression model\cite{Aerts2018}, written as:

\begin{equation}
\label{regression}
Y_{ji} = \boldsymbol{x}_{ji}^\top \boldsymbol{\beta}_j,
\end{equation}

\noindent where $Y_{ji}$ corresponds to the observable $i$ of the grid point $j$ (e.g. $\Delta P_{1, 1, 80, j}$), while $\boldsymbol{x}_{ji}$ are functions of $\boldsymbol{\theta}$ based on the principle of fractional polynomials\cite{Aerts2018}, and $\boldsymbol{\beta}_j$ are the regression coefficients. For each observable $Y_{ji}$, the optimal number of regression coefficients is determined from statistical model selection based on the Bayesian Information Criterion applied to the nested regression models\cite{Claeskens2008}. The typical number of regression coefficients to approximate each period spacing prediction ranges from 18 to 30, depending on the grid considered.

Theoretical period spacing patterns ($\Delta P^{\rm theo}$) covering ranges in radial orders from $n \in [1, 80]$ were matched to observed values ($\Delta P^{\rm obs}$) in three different ways: 1) the theoretical patterns are built starting from the lowest mode period in the observed patterns, matching it by finding the $\Delta P^{\rm theo}$ value that results in the smallest difference in any given grid point and assigning the rest of the theoretical $\Delta P$ values such that they are consecutive in radial order; 2) the matching of the theoretical patterns is done from the two period spacings resulting from the observed mode with the highest amplitude in the periodograms and enforcing consecutive radial orders; 3) among the differences between the $\Delta P^{\rm obs}$ and all of the $\Delta P^{\rm theo}$ in a given grid point, we search those delivering the longest matching sequence in consecutive radial orders and assign the rest of the $\Delta P^{\rm theo}$ values according to this sequence by enforcing consecutive radial orders for the remaining $\Delta P^{\rm theo}$. For each of the three ways of constructing $\Delta P^{\rm obs}$ and $\Delta P^{\rm theo}$, we search for the best fit between the observables and theoretical predictions by applying the statistical method based on the Mahalanobis distance (MD) as the merit function\cite{Aerts2018}. The details are omitted here for brevity. This merit function represents a more general distance compared to the Euclidean distance, which is a special case and corresponds to a $\chi^2$ based merit function. The Mahalanobis distance optimization takes the form 

\begin{equation}
\label{maldist}
{\rm MD} =\arg\!\min_{{\boldsymbol{\theta}}} \left\{ (\Delta P^{\rm theo}({\boldsymbol{\theta}}) - \Delta
  P^{\rm obs})^\top (V + \Sigma)^{-1} 
(\Delta P^{\rm theo}({\boldsymbol{\theta}}) - \Delta P^{\rm obs}) \right\},
\end{equation}

\noindent where the notation $X^\top$ stands for the transpose of $X$, $V$ is the variance-covariance matrix of the vector $\Delta P^{\rm theo} ({\boldsymbol{\theta}})$ for each of the grids $\boldsymbol{\psi}_1, \ldots, \boldsymbol{\psi}_8$ and $\Sigma$ is the matrix with diagonal elements given by the observational errors of the $i=1, \ldots, N$ measured period spacings. The values of $\Delta P^{\rm theo}({\boldsymbol{\theta}})$ are taken from the statistical grids of stellar model predictions constructed for each of the eight $\boldsymbol{\psi}$ values. The Mahalanobis distance defined by Eq.\,(\ref{maldist}) takes full account of the fact that theoretical uncertainties in the predictions $\Delta P^{\rm theo}$ are typically two orders of magnitude larger than the observational uncertainties of $\Delta P^{\rm obs}$ measured from 4-year {\it Kepler\/} data \cite{Aerts2018} and includes the overall correlated nature of the parameters ${\boldsymbol{\theta}}$ and the observables $\Delta P^{\rm theo}$. The stability of the solution for MD resulting from Eq.\,(\ref{maldist}) is determined by the eigenvalues of the matrix $V + \Sigma$, which captures the combined theoretical and measurement covariance structure of the quantities used in the modelling. This stability is set by the so-called condition number of this matrix. We retain the solution of Eq.\,(\ref{maldist}) for that problem set among the three ways of constructing the theoretical period spacing patterns delivering the smallest condition number.

The solution for MD is computed several times for each star and for each grid $\boldsymbol{\psi}_1, \ldots, \boldsymbol{\psi}_8$, relying on different combinations of $T_{\rm eff}$, $\log\,g$, and $\log\,(L/L_\odot)$ as error box to compute the statistical regression models for $\Delta P^{\rm theo}({\boldsymbol{\theta}})$. That is because the measurement quality of these classical observables is different per star. Moreover, the character of the period spacing patterns also differs strongly from star to star. Some stars deliver more than one option to construct the observational patterns \cite{Szewczuk2018}. For all these various solutions from Eq.\,(\ref{maldist}), we kept the one relying on the variance-covariance matrix with the lowest condition number. In practice, the condition numbers encountered for each of the 26 SPB stars range over 26 - 7000.

The retained solutions to Eq.\,(\ref{maldist}) were then subjected to statistical model selection based on the Aikaike Information Criterion corrected for small numbers \cite{Claeskens2008}, AICc, given that the number of components in the observed patterns for the 26 SPB stars range from 7 to 34 (cf. Supplementary Figures\,1 to 26).  The overall best solution for ${\boldsymbol{\theta}}$ per star and per grid $\boldsymbol{\psi}_1, \ldots, \boldsymbol{\psi}_8$ is then selected from the lowest AICc combined with visual inspection among all the computed cases for the MD and AICc values. A fully automated process is sub-optimal because the AICc, as well as any alternative selection criterion, depends on outlier behaviour\cite{Claeskens2008}. For our application to period spacing patterns, this implies that the AICc can pick a solution with a low AICc value due to a particular trapped mode, which may act as an outlier in the pattern. The diversity in deviations from a constant period spacing is large, as illustrated in Supplementary Figures\,1 to 26, such that visual inspection is warranted. The outcome of the asteroseismic modelling for the 26 SPB stars is listed in Supplementary Tables\,1 and 2, and shown graphically in the bottom panels of Supplementary Figures\,1 to 26. 

For the global stellar ${\boldsymbol{\theta}}$ components, i.e., the mass, metallicity, evolutionary stage, and rotation frequency, it is meaningful to compute an averaged value weighted according to MD across the eight grids. The standard deviation with respect to this average provides an estimate of the systematic uncertainty for these parameters due to the unknown internal mixing physics. The outcome is provided in Supplementary Table~2 and shows that the four ${\boldsymbol{\theta}}$ components agree with the averaged value across the eight grids to within the standard deviation for 19 of the 26 SPB stars. Precision estimation of ${\boldsymbol{\theta}}$ per grid is a notoriously difficult issue because the components of the parameter vector are strongly correlated for g-mode asteroseismology in stars with a convective core \cite{Moravveji2015}. It is not meaningful to assess the precision of each ${\boldsymbol{\theta}}$ ignoring this correlation. Rather, one has to compute 6-D uncertainty regions. We handle this by a Monte Carlo approach, by perturbing the regression coefficients $\boldsymbol{\beta}$ in Eq.\,(\ref{regression}) 100 times and recomputing the MD and ${\boldsymbol{\theta}}$ solution accordingly. The outcome is shown graphically by means of projected error ranges in Supplementary Figures\,1 to 26, but we stress that these projections onto a single ${\boldsymbol{\theta}}$ component axis should not be interpreted as being independent error estimates in 1-D.


\section*{Data availability}
The data that support the plots within this paper and other findings of this study are available from the corresponding author upon reasonable request.

\section*{Code availability}
The iterative pre-whitening code is freely available and documented at\\
{\tt\small https://github.com/IvS-KULeuven/IvSPythonRepository}. The stellar evolution code, MESA, is freely available and documented at {\tt\small http://mesa.sourceforge.net/}. The stellar pulsation code, GYRE, is freely available and documented at\\ {\tt\small https://bitbucket.org/$\sim$rhdtownsend/gyre/wiki/Home}.

\section*{Acknowledgements}
The authors are grateful to the MESA and GYRE code developers for their efforts, public dissemination, and training initiatives to make their software so accessible to the worldwide astrophysics community. The authors thank Dr.\ Sylvia Ekstr{\"o}m of the Geneva Observatory for having provided mixing profiles from Georgy et al. (2013) in electronic format. We acknowledge the work of the teams behind the NASA {\it Kepler\/} and ESA Gaia space missions. 
This work is based on observations with the HERMES spectrograph at the Mercator Telescope which is operated at La Palma/Spain by the Flemish Community. This research has made use of the SIMBAD database, operated at CDS, Strasbourg, France and of NASA’s Astrophysics Data System. The research leading to these results has received funding from the European Research Council (ERC) under the European Union’s Horizon 2020 research and innovation programme (grant  agreement N$^\circ$670519: MAMSIE), from
the National Science Foundation (Grant No. NSF PHY-1748958),
from the KU\,Leuven Research Council (grant C16/18/005: PARADISE), and from the Research Foundation Flanders (FWO) by means of PhD Fellowships to M.M. and S.G. under contract N$^\circ$11F7120N and N$^\circ$11E5620N, and a senior post-doctoral fellowship to D.M.B. with grant agreement N$^\circ$1286521N. Funding for the {\it Kepler\/} Mission was provided by NASA's Science Mission Directorate. Gaia data are being processed by the Gaia Data Processing and Analysis Consortium (DPAC); funding for the DPAC is provided by national institutions, in particular the institutions participating in the Gaia MultiLateral Agreement (MLA). 


\section*{Author contributions}
M.G.P performed frequency analysis and mode identification, wrote code to include mixing profiles in MESA, computed asteroseismic observables, implemented and applied the modelling procedures, interpreted the results, and wrote part of the text. C.A. defined the research, developed the modelling procedure, interpreted the results, and wrote part of the text. P.I.P constructed light curves from the raw {\it Kepler\/} data and discovered the targets to be new SPB stars. M.M. wrote code to include mixing profiles in MESA and assessed the capacity of observables used for the modelling. S.G. determined abundances from spectroscopy. T.M.R. computed and provided envelope mixing profiles due to internal gravity waves. G.M. provided advice on the parameter estimation and statistical model selection and performed the cluster analysis. 
S.B., S.G., and D.M.B. contributed to the frequency analysis and interpretation. All authors contributed to the discussions and have read and iterated upon the text of the final manuscript. 

\section*{Competing interests}
The authors declare no competing interests. 

\section*{Materials \& Correspondence}
Correspondence and requests for materials should be addressed to the corresponding author May Gade Pedersen (mgpedersen@kitp.ucsb.edu). 

\section*{Additional information}

\textbf{Supplementary Information} is available for this paper.

\noindent\textbf{Correspondence and requests for materials} should be addressed to M.G.P.

\noindent\textbf{Reprints and permissions information} is available at {\tt\small www.nature.com/reprints}

\end{document}